\documentclass{article}
\newcounter{todocounter}
\newcommand{\todonum}{\stepcounter{todocounter}{(\thetodocounter)}}

\usepackage{fullpage,color}
\usepackage{amsmath,amsthm}

\def\shownotes{1}   
\ifnum\shownotes=1
\newcommand{\authnote}[2]{{ $<<$\textsf{\footnotesize \todonum\  #1 notes:  #2}$>>$}}
\else
\newcommand{\authnote}[2]{}
\fi

\mathchardef\mhyphen="2D 

\title{Revisiting Fast Practical Byzantine Fault Tolerance}

\author{Ittai Abraham, Guy Gueta, Dahlia Malkhi\\
VMware Research\\
\mbox{}\\
\mbox{}\\
with:\\
Lorenzo Alvisi (Cornell),\\
Rama Kotla (Amazon),\\
Jean-Philippe Martin (Verily)
}

\begin{document}

\maketitle


\newcommand{\ignore}[1]{}

\begin{abstract}
In this note, we observe a safety violation in
Zyzzyva~\cite{K07,K08,KAD09} and a liveness violation in FaB~\cite{MA05,MA06}. 
To demonstrate these issues, we require relatively simple scenarios,
involving only four replicas, and one or two view changes. 
In all of them, the problem is manifested already in the first log slot. 
\end{abstract}

\section{Introduction}
A landmark solution in achieving replication with Byzantine fault tolerance has been the Practical Byzantine Fault Tolerance (PBFT) work by Castro and Liskov~\cite{CL99,CL02}. 
Since the PBFT publication, there has been a stream of works aiming to improve the efficiency of PBFT protocols.   
One strand of these works revolves around \textit{optimism}~\cite{K02,MA05,MA06,K07,K08,KAD09,upright,sevenBFT}. In this strand, the
focus is on providing a \textit{fast} common case (i.e., when there
are no link or server failures). In other cases, optimistic solutions fall back
to some backup implementation with strong progress guarantees.

In this note, we observe that several key works in the ``optimistic strand'' do
not deal with optimism correctly. 
In particular, we first present in \S\ref{chap:ZZ} safety violations in
Zyzzyva~\cite{K07,K08,KAD09}. We then demonstrate in \S\ref{chap:fab} how being
``overly safe'' gets FaB~\cite{MA05,MA06} stuck. 
To demonstrate these issues, we require relatively simple scenarios,
involving only four replicas, and one or two view changes. 
In all of them, the problem is manifested already in the first log slot. 

We also briefly observe below that in other fast Byzantine replication solutions, an optimistic track is not fully intertwined with a regular protocol, hence they are \textit{less fast}. 

\medskip

It therefore appears that the challenge posed in~\cite{fastpaxos} of providing \textit{Byzantine Fast Paxos} is left open:
\begin{quote}
``Fast Paxos can also be generalized to a Fast Byzantine Paxos algorithm that requires only two message delays between proposal and learning in the absence of collisions. (However, a single malicious proposer can by itself create a collision.)''~\cite{fastpaxos}
\end{quote}

That is, none of the fast Byzantine agreement works we are aware of provides 
a solution that simultaneously addresses (i) optimal step-complexity, (ii) optimal resilience, (iii) safety against failures of less than a third of the system, and (iv) progress during periods of partial synchrony.

Our team has worked out a full solution, and will publish a follow up to this report in the near future.


\subsection*{Preliminaries}

The focus of this work is providing state-machine-replication (SMR) 
for $n$ \textit{replicas}, $f$ of which can be Byzantine faulty. An unbounded
set of \textit{clients} may form \textit{requests} and submit them to replicas.
We refer to members of the system, replicas or clients, as \textit{nodes}. 
The communication among nodes is authenticated, reliable, but asynchronous; that is, we assume that a message sent
from a correct node to another correct node is signed and eventually arrives. 

At the core of SMR is a protocol for deciding on a growing log of operation requests by clients, satisfying the following properties: 
%

\begin{description}

\item[Agreement] If two correct replicas commit decisions at log position
 $s$, then the decisions are the same.

\item[Validity] If a correct replica commits a decision at some log position,
then it was requested (and signed) by some client.

\item[Liveness] If some correct client submits a request, and the system is
eventually partially-synchronous~\cite{DLS_jacm88}, then eventually the replicas
commit some decision.

\end{description}

\subsection*{View Change}

The solutions we discuss employ a classical framework that revolves around an explicit ranking among proposals via \emph{view} numbers.

Replicas all start with an initial view, and progress from one view to the next.
They accept requests and respond to messages only in their current view.

In each view there is a single designated \textit{leader}.  
In a view, zero or more decisions may be reached.
This strategy separates safety from liveness: It maintains safety even if the system exhibits arbitrary communication delays and again up to $f$ Byzantine failures; it provides progress during periods of synchrony. 

If a sufficient number of replicas suspect that the leader is faulty, then a
view change occurs and a new leader is elected.
The mechanism to trigger moving to a higher view is of no significance for
safety, but it is crucial for liveness. On the one hand, replicas must not be
stuck in a view without progress; on the other hand, they must not move to a
higher view capriciously, preventing any view from making progress. 
Hence, a replica moves to a higher view if either
a local timer expires, or if it receives new view suggestions from
$f+1$ replicas. Liveness relies on having a constant
fraction of the views with a correct leader, whose communication with correct replicas is timely, 
thus preventing $f+1$ replicas from expiring.

Dealing with leader replacement is the pinnacle of both safety and liveness. 
A core aspect in forming agreement against failures is the need for new leaders
to safely adopt previous leader values. The reason is simple, it could be that a
previous leader has committed a decision, so the only safe thing to do is adopt
his value. 

In the prevailing solutions for the benign settings (DLS~\cite{DLS_jacm88}, Paxos~\cite{paxos}, VR~\cite{oki_podc88}, Raft~\cite{ongaro_atc14}), leader replacement is done by reading from a quorum of $n-f$ replicas and choosing the value with the maximal view \footnote{In DLS, the term \textit{phase} is used, and in Paxos, \textit{ballot}.} number.
Note that
$n-f$ captures a requirement that the quorum intersects every leader quorum in previous views (not only the most recent one). It is crucial to take into consideration how leader quorums of multiple previous views interplay.
Choosing the value with the maximal view number is crucial because there may be multiple conflicting values and choosing an arbitrarily value is not always a safe decision.


A similar paradigm holds in PBFT~\cite{CL99,CL02}. The new leader needs to read from a quorum of $n-f$
replicas and choose a value with the maximal view number.
Different from the benign case, in
the Byzantine settings, uniqueness is achieved by using enlarged, Byzantine
quorums~\cite{MR_dc98}. Byzantine quorums guarantee intersection not just in any
node but in a correct node. 

In Byzantine settings, a correct node also needs to prove a decision value to a new leader.
This is done in PBFT\footnote{We refer here to the PBFT version with signed messages~\cite{CL99}.} by adding another phase before a decision.  
The first phase ensure uniqueness via \emph{prepare} messages from $n-f$ nodes.
In the second phase, nodes send a \textit{commit-certificate} consisting of $n-f$ prepare messages. A decision can be reached when $n-f$ nodes have sent a commit-certificate.  
The two-phase scheme guarantees that if there is a decision, there is a correct node that passes a
commit-certificate to the next view. 

\subsection*{Sacrificing Resilience}
The extra PBFT phase may be avoided by somewhat sacrificing resilience and
using $n=5f+1$, as in FaB~\cite{MA05,MA06}, Zyzzyva5~\cite{K07}, and Q/U~\cite{QU}. Here, the
intersection between a potential decision quorum and a view-change quorum has
$2f+1$ correct nodes, enough to provide both uniqueness and transfer of value.

\subsection*{Kursawe's Solution}
Addressing a much more limited scope,
Kursawe provided in 2002 a simple black box technique to transform any Asynchronous Byzantine Agreement (ABA) protocol (with a sufficiently strong validity property) into a consensus protocol that has an optimistic fast path~\cite{K02}. 
It works as follows.
 
 
  

There are two possible commit tracks, and they may be combined (some nodes commit in the fast, some not).
In the fast track, a node decides if all nodes prepare an identical value.
In the fall-back track, any Byzantine agreement protocol is invoked,
where nodes use their prepare values as initial inputs. The only requirement
from the agreement protocol is that it satisfies the following validity property:
 
\begin{description}
      \item[Byzantine validity:] If all correct nodes start with the same
input $v$,
then the decision must be $v$.
\end{description}
 
This succinct solution framework is (almost trivially) correct. 
However, the recovery stage
does not utilize the prepare steps which were already performed in the fast track. Hence, whereas the fast track is fast, the fall-back track is not optimal.

Additionally, as we already noted, it addresses a problem of a much more limited scope: It solves only a single-shot consensus; it
does not address state replication (execution) at all. 

\subsection*{FaB}
FaB~\cite{MA05,MA06} extends Kursawe's solution in several ways. First, the
prepare messages from the fast track are input to the recovery phase, thus
reducing the number of steps in recovery mode. In this way, the FaB recovery
mode has the same overall cost as standard PBFT. Second, FaB extends the treatment to
a parameterized failure model of $n=3f+2t+1$. Thus, by appropriately increasing
the system size, fast termination is achieved despite up to $t$ non-leader
Byzantine failures, whereas safety is guaranteed against $f$. 

To achieve these enhancements, FaB cannot employ a Byzantine agreement protocol for recovery as a ``black-box''.  Unfortunately, opening the recovery agreement protocol and incorporating the consensus steps into the FaB framework resulted in the omission we surface here (see \S\ref{chap:fab}).

\subsection*{Zyzzyva}
Zyzzyva borrows from FaB the method for efficiently intertwining the optimistic
fast track with the recovery track. It enhances the approach in a number of
dimensions. Zyzzyva provides a state replication protocol, whereas FaB is a
single shot consensus solution. Zyzzyva employs speculation in the execution of
state updates, allowing a high throughput pipeline of state-machine replication,
which is out of the FaB scope. Finally, a new leader in Zyzzyva cannot get
``stuck'' choosing a safe value as in FaB (\S\ref{chap:fab}).
Unfortunately, the view-change protocol in Zyzzyva
fails to provide safety against a faulty leader, as described
in~\S\ref{chap:ZZ}. 

\subsection*{Upright}
The Zyzzyva view-change protocol has been employed in UpRight ~\cite{upright}, which also incorporates the parameterized failure model of $n=3f+2t+1$ from FaB. The goal of UpRight is to build an engineering-strength BFT engine. The UpRight paper does not provide a full description of the algorithm, and rather indicates that it adopts these two previous solutions. 


\subsection*{The Next 700 BFT Protocols}
In \textit{The Next 700 BFT Protocols}, Aublin et al.~\cite{sevenBFT} provide a principled approach to view-change in BFT protocols. Their approach switches not only leaders, but also entire regimes, in order to respond to adaptive system conditions. One node of the 700 BFT protocol family is AZyzzyva, a protocol that combines the speculative (fast) path of Zyzzyva in a protocol called Zlight with a recovery protocol, e.g., PBFT. If Zlight fails to make progress, it switches to a new view that executes PBFT for a fixed number $k$ of log slots. In this sense, AZyzzyva falls back to the approach of Kursawe~\cite{K02}, while extending it to a pipeline of state-machine commands and implementing a replicated state-machine. Indeed, Azyzzyva is simple and principled, and it is not vulnerable to the safety violations of Zyzzyva exposed here (\S\ref{chap:ZZ}). At the same time, the Azyzzyva recovery path requires more steps than the two-phase protocol of Zyzzyva. Additionally, Azyzzyva requires to wait for a commit decision (of $k$ slots) to switch back from PBFT to Zlight.

\newpage
\section{Revisiting the Zyzzyva View-Change} \label{chap:ZZ}
\subsection{Introduction}

The Zyzzyva~\cite{K07,K08,KAD09} has two commit paths. A two-phase path that resembles PBFT and a fast path. 

The fast path does not have commit messages, and a client commits a decision by
seeing $3f+1$ prepare messages\footnote{Note that, the terms \textit{prepare}
and \textit{commit} are taken from PBFT; In Zyzzyva, the leader proposal message
is called {\tt ORDER-REQ} and the acknowledgements by replicas which are akin to
prepare messages are called {\tt SPEC-RESPONSE}.}. 
The optimistic mode is coupled with a recovery mode that guarantees progress in face of failures. The recovery mode intertwines the PBFT two-phase steps into the protocol.

Quoting from ~\cite{KAD09}, "Fast agreement and speculative execution have profound effects on Zyzzyva’s view change subprotocol."

Indeed, in Zyzzyva, a possible decision value is transferred across views in two possible ways, corresponding to the two decision tracks of the protocol (fast and two-phase):
In the fast track, a possible decision value manifests itself as $f+1$ prepare
messages. 
In the two-phase track, it manifests itself as a commit-certificate (as in PBFT). Combining the two, Zyzzyva prefers a commit-certificate over $f+1$ prepares; and among two commit-certificates, it prefers the one with the longer request-log.  

Here we show that either one of these rules may lead to violating safety.

The omissions are quite subtle, because unless a leader equivocates, a commit-certificate will not conflict with fast-paths of higher views.

Likewise, unless a leader equivocates, the log can only grow from one view to the next. Hence, in benign executions, higher views have longer (or at least non-decreasing\footnote{it seems that another, minor omission in the Zyzzyva protocol is that it does not explicitly indicate how to break ties in case of two maximal commit-certificates, of same length}) sequence of commands, and the notions of highest view and longest request-log will be the same. 

Nevertheless, 
we show that both these strategies do not provide safety, and permit the scenarios we surface here, where Zyzzyva breaks safety.

\subsection{A Skeletal Overview of Zyzzyva}

We start with an overview of Zyzzyva. 
Our description is merely skeletal, and 
%
glosses over many engineering details: 
We assume that all messages are signed and are forwarded carrying their signatures;
we neglect the mechanism for checkpoint and space reclamation; 
and we do not optimize for messages sizes and crypto operations. These details and
optimizations are covered in the Zyzzyva paper, and are omitted here for brevity
and clarity.

As in the original paper, we break the  Zyzzyva agreement protocol into three
sub-protocols, a fast-track sub-protocol, a two-phase sub-protocol, and a
view-change sub-protocol. 

\paragraph{Messages.}

Since we mostly adopt the notation and terminology from PBFT, we start with a quick
reference guide, mapping Zyzzyva's message types to PBFT's. 


\begin{description}

\item{Client-request:}
A \textit{client-request} (REQUEST) from a client to the leader contains some
operation $o$, whose semantics are completely opaque for the purpose of this
discussion. 

\item{Ordering-request:}
A leader's \emph{pre-prepare} message is called an \textit{ordering-request}
(ORDER-REQ), and contains a leader's
log of client requests $OR_n = (o_1, ..., o_n)$. 
(In practice, the leader sends only the last request and a hash of the history of prior 
operations; a node can request the leader to re-send any missing 
operations.)

\item{Ordering-response:}
When a replica \textit{accepts} a valid pre-prepare request, it speculatively
executes it and sends the result in 
a \emph{prepare} message called an \emph{ordering-response} (SPEC-RESPONSE). 

\item{Commit-request:}
A \textit{commit-request} (COMMIT) from the client to the replicas includes a \textit{commit-certificate} $CC$, 
a set of $2f+1$ signed replica responses (SPEC-RESPONSE) to an
(identical) ordering-request $OR_n$.

\item{Commit-response:}
When a replica obtains a valid commit-certificate $CC$ for $OR_n$, it responds
to client requests in $OR_n$
with a \textit{commit} message called a \emph{commit-response} (LOCAL-COMMIT).

\item{View-change:}
A \textit{view-change} (VIEW-CHANGE) message from a replica to the leader of a
new view captures the replica's \emph{local state}. 

\item{New-view:}
A \textit{new-view} (NEW-VIEW) message from the leader of a new view contains a
set $P$ of view-change messages the leader collected, which serves as a new-view
proof. It includes a new ordering request 
$G_n = (o_1, ..., o_n)$. 

\end{description}

\paragraph{The fast-track sub-protocol.}

Zyzzyva contains a fast-track protocol in which a client learns the result of a
request in
only three message latencies, and only a linear number of crypto operations. 
It works as follows. 

%

A client sends a request $o$ to the current leader. The
current leader extends its local log with the request $o$ to $OR_n$, and sends a
pre-prepare (ordering-request) carrying $OR_n$. We did not say how a leader's local
log is initialized. Below we discuss the protocol for a leader to pick an
initial log when starting a new view. 

A replica \textit{accepts} a pre-prepare from the leader of the current
view if it has valid format, and it extends
any previous pre-prepare from this leader. 
Upon accepting a pre-prepare, a replica extends its local log to $OR_n$ 
It speculatively executes it, 
and sends the result directly to the client in a \emph{prepare} message.

A decision is reached on $OR_n$ in view $v$ in the fast track 
when $3f+1$ distinct replicas have sent a prepare message for it. 

\paragraph{The two-phase sub-protocol.}

If progress is stalled, then a client waits to collect a
\emph{commit-certificate}, a set of $2f+1$ prepare responses for 
$OR_n$. 
Then the client sends a commit-request carrying the commit-certificate to the
replicas. A replica responds to a valid commit-request with a
\emph{commit} message.

A decision is reached on $OR_n$ in view $v$ in the two-phase
track when $2f+1$ distinct replica have sent a commit message for it.

\paragraph{The view-change protocol.}

The core mechanism in Zyzzyva for transferring safe values across views is
for a new Zyzzyva leader to collect a set $P$ of
view-change messages from a quorum of $2f+1$ replicas. Each replica sends a
view-change message containing the replica's \textit{local state}:
Its local request-log,
and the commit-certificate with the highest view number it responded to with a
commit message, if any.

The leader processes the set $P$ as follows.

\begin{enumerate}
 \item
  Initially, it sets a base log $G$ to an empty log.
  
  \item
  If any view-change message contains a valid commit-certificate, then it
selects the one with the longest request-log $OR_n$ and copies $OR_n$ to $G$.  
 
\item
If $f+1$ view-change messages contain the same request-log $OR'_m$,
   then it extends the tail of $G$ with requests from $OR'_m$. (If there are two
$OR'_m$ logs satisfying this, one is selected arbitrarily.)
   
\item
Finally, 
it pads $G$ with null request entries up to the length of the longest log of any
valid prepare.

\end{enumerate}

The leader sends a new-view message to all the replica. The message
includes the new view number $v+1$, the set $P$ of view-change messages the
leader collected as a proof for new-view $(v+1)$, and a request-log 
$G$. A replica accepts a new-view
message if it is valid, and \emph{adopts} the leader log. It may need to roll
back speculatively executed requests, and process new ones.

\newcommand{\begindetail}{\hrulefill \textit{Additional details.}}
\newcommand{\detailoff}{\hrulefill \textit{To here.}}

\newenvironment{detail}[1]
{
\ifx\addldetail\undefined
undefed
\else
  \if\addldetail1
  \begindetail
 #1
  \else
  \fi
\fi
}
{
\ifx\addldetail\undefined
undefed
\else
  \if\addldetail1
	\detailoff
  \else
  \fi
\fi
}

\subsection{Breaking Safety: First Scenario} \label{sec:break}

We now proceed to demonstrate that the view-change mechanism in Zyzzyva does not guarantee safety. The overview of Zyzzyva we provided above should suffice to understand the scenarios below; for precise detail and notation of the Zyzzyva protocol, the reader is referred to ~\cite{K08}.

Our first scenario demonstrates 
that the criterion for combining fast-track decision with two-phase decision may lead to a safety violation. 
In particular, prioritizing commit-certificate over $f+1$ prepares, as done in Zyzzyva, is not always correct.

Our scenario requires four replicas $i_1$, $i_2$, $i_3$, $i_4$, of which one, $i_1$, is Byzantine. 
It proceeds in $3$ views, and arrives at a conflicting decision on the first log position.

\def\addldetail{0}
\paragraph{View 1: Creating a commit-certificate for $(a)$.}

\begin{enumerate}
\item Two clients $c_1$, $c_2$ provide a leader $i_1$ of view $1$ with well-formed requests (${\tt REQUEST}$) $a$ and $b$, respectively.

\item In view $1$, the leader $i_1$ sends to replicas $i_2$ and $i_3$ a
pre-prepare (${\tt ORDER\mhyphen REQ}$) for $a$. 

\item The leader $i_1$ (Byzantine) equivocates and sends replica $i_4$ a
conflicting pre-prepare for $b$.

\item
Replicas $i_2$ and $i_3$ accept the leader's well-formed pre-prepare, and
speculatively execute $a$. They obtain a speculative result and send it in a
prepare response (${\tt SPEC\mhyphen RESPONSE}$) to $c_1$.

\item Client $c_1$ collects prepares from $i_1$, $i_2$ and $i_3$
for the request-log $(a)$. These responses constitute a commit-certificate,
denoted $cert$. 

Then the client expires waiting for additional responses. It
sends a commit-request (${\tt COMMIT}$) for $(a)$ that includes the
commit-certificate $cert$. The commit-request reaches only $i_1$.

\end{enumerate}

\paragraph{View 2: Deciding $(b)$.}

\begin{enumerate}

\item All further messages are delayed, forcing the system to go through a view change.

\item In view $2$, the leader $i_2$ collects view-change messages (${\tt VIEW\mhyphen CHANGE}$) from itself, from $i_1$ and from $i_4$ as follow: 
  \begin{itemize}
  \item Replica $i_2$ sends its local log $(a)$.
 
  \item Replica $i_4$ sends its local log $(b)$:

  \item Replica $i_1$ (which is Byzantine) joins $i_4$ and sends a request-log $(b)$. 

  \end{itemize}  
  
Based on these view-change messages, $i_2$ constructs a 
new request-log $G$ consisting of $(b)$, and sends it in a new-view message (${\tt NEW\mhyphen VIEW}$) 
to replicas.

\item
Every replica accepts the leader $i_2$ well-formed new-view message. Upon
accepting it, each replica zeros its local log (undoing $a$, if needed).
All replicas adopt the leader request-log $(b)$ and speculatively execute $b$. 
They obtain a speculative result and send it in a
response (${\tt SPEC\mhyphen RESPONSE}$) to $c_2$.

\item 
The client $c_2$ of $b$ collects speculative-responses from all replicas, and 
\textbf{$b$ becomes successfully committed at log position $1$.}

\end{enumerate}

\paragraph{View 3: Choosing the wrong commit-certificate.}

\begin{enumerate}

\item All further messages are delayed, forcing the system to go through a view change.

\item In view $3$, the leader $i_3$ collects view-change messages (${\tt VIEW\mhyphen CHANGE}$) from itself, from $i_1$ and from $i_4$ as follow: 
  \begin{itemize}
  \item Replica $i_1$, which is Byzantine, hides the value it prepared in view $2$, and sends commit-certificate $cert$ (see above) for $(a)$.

  \item Replicas $i_3$ and $i_4$ send their local logs $(b)$.

  \end{itemize}
    
Based on these view-change messages, $i_3$ chooses $cert$, the
commit-certificate, and adopts it. It constructs a new request-log $G$ 
consisting of requests $(a)$, 
and sends it in a new-view message  (${\tt NEW\mhyphen VIEW}$) to replicas.

\item
Each replica accepts the leader $i_3$ well-formed new-view message. Upon
accepting it, replicas \textbf{zero their local logs, undoing $\mathbf{b}$ as needed.}
Then they speculatively execute $a$, send the result, and 
\textbf{$a$ becomes
successfully committed at log position 1.}

\end{enumerate}

\subsection{Breaking Safety: Second Scenario}

The second scenario demonstrates that the criterion for combining two-phase decisions from different views may lead to a safety violation. In particular, prioritizing the longest commit-certificate, as done in Zyzzyva, is not always correct.

Our second scenario again requires four replicas $i_1$, $i_2$, $i_3$, $i_4$, of which one, $i_1$, is Byzantine. 
It proceeds in $3$ views, and arrives at a conflicting decision on the first log position. In order to construct commit-certificates of different lengths, it utilizes four operation requests, $a_1$ by client $c_1$, $a_2$ by $c_2$, $b_1$ by $c_3$, and $b_2$ by $c_4$. 

%

\def\addldetail{0}
\paragraph{View 1: Creating a commit-certificate for $(a_1, a_2)$.}

\begin{enumerate}
\item Four clients $c_1$, ..., $c_4$ provide a leader $i_1$ of view $1$ with well-formed requests (${\tt REQUEST}$) for $a_1$, $a_2$, $b_1$, and $b_2$, respectively.

\item In view $1$, the leader $i_1$ sends to replicas $i_2$ and $i_3$ two
pre-prepare messages (${\tt ORDER\mhyphen REQ}$). The first 
one is for $a_1$ at log position $1$. The second one is for $a_2$ at log position $2$, succeeding $a_1$. 

\item The leader $i_1$ (Byzantine) equivocates and sends replica $i_4$ two
conflicting pre-prepare requests. The first 
one is for $b_1$ at log position $1$. The second one is for $b_2$ at log position $2$ succeeding $b_1$.

\item
Replicas $i_2$ and $i_3$ accept the relevant leader's well-formed pre-prepares,
and speculatively execute $a_1$ followed by $a_2$. They obtain speculative
results and send each result in a corresponding prepare response (${\tt SPEC\mhyphen RESPONSE}$) to its requesting client.

\item The client $c_2$ of $a_2$ collects prepares from $i_1$, $i_2$ and $i_3$
for the request-log $(a_1, a_2)$. These responses constitute a
commit-certificate, denoted $cert_1$. 

Then the client expires waiting for
additional responses. It sends a commit-request (${\tt COMMIT}$) for $(a_1,a_2)$
that includes the commit-certificate $cert_1$. The commit-request reaches only
$i_3$.

\end{enumerate}

\paragraph{View 2: Deciding $(b_1)$.}

\begin{enumerate}

\item All further messages are delayed, forcing the system to go through a view change.

\item In view $2$, the leader $i_2$ collects view-change messages (${\tt VIEW\mhyphen CHANGE}$) from itself, from $i_1$ and from $i_4$ as follow: 
  \begin{itemize}
  \item Replica $i_2$ sends its local log $(a_1,a_2)$.
 
  \item Replica $i_4$ sends its local log $(b_1, b_2)$.

  \item Replica $i_1$ (which is Byzantine) joins $i_4$ and sends a request-log $(b_1,b_2)$. 
      
  \end{itemize}  
  
Based on these view-change messages, $i_2$ constructs a 
new request-log $G$ consisting of $(b_1, b_2)$, and sends it 
in a new-view message (${\tt NEW\mhyphen VIEW}$) to replicas.

\item
Each replica among $i_1$, $i_2$ and $i_4$ accepts the leader $i_2$ well-formed
new-view message. Upon accepting it, replica $i$ zeros its local log (undoing
$a_1, a_2$ as needed), and adopts the leader request-log $(b_1, b_2)$. It
first proceeds to speculatively execute $b_1$,
obtains a speculative result, and sends it in in a response (${\tt
SPEC\mhyphen RESPONSE}$) to $c_3$.

\item 
The client $c_3$ of $b_1$ collects speculative-responses from $i_1$, $i_2$ and
$i_4$ for the request-log $(b_1)$. These responses constitute a
commit-certificate, denoted $cert_2$. 

Then the client expires waiting for
additional responses. It sends a commit-request (${\tt COMMIT}$) for $(b_1)$
that includes the commit-certificate $cert_2$. 

\item
Upon receiving the well-formed commit-request, replicas $i_1$, $i_2$, and $i_4$ respond to client $c_3$ with a commit message (${\tt LOCAL\mhyphen COMMIT}$).

\item 
The client collects these commit messages and \textbf{$\mathbf{b_1}$ becomes successfully
committed at log position $\mathbf{1}$}.

\end{enumerate}

\paragraph{View 3: Choosing the wrong, maximal commit-certificate.}

\begin{enumerate}

\item All further messages are delayed, forcing the system to go through a view change.

\item In view $3$, the leader $i_3$ collects view-change messages (${\tt VIEW\mhyphen CHANGE}$) from itself, from $i_1$ and from $i_4$ as follow: 
  \begin{itemize}
  \item Replica $i_3$ sends commit-certificate $cert_1$ (see above) for $(a_1, a_2)$.

  \item Replica $i_4$ sends commit-certificate $cert_2$ (see above) for $(b_1)$,
and its local log $(b_1, b_2)$.

  \item Replica $i_1$ (Byzantine) can join either one, or even send an
view-change message with an empty log. 
  \end{itemize}
    
Based on these view-change messages, $i_3$ chooses $cert_1$, the
commit-certificate with the longest request-log, and adopts it. It constructs a 
new request-log $G$ consisting of $(a_1, a_2)$, and sends it 
in a new-view message  (${\tt NEW\mhyphen VIEW}$) to replicas. 

\item
Each replica accepts the leader $i_3$ well-formed new-view message. Upon accepting it, 
replicas \textbf{zero their local logs, undoing $\mathbf{b_1}$ as needed.}
Then they speculatively execute $a_1$, send the result, and 
\textbf{$\mathbf{a_1}$ becomes
successfully committed at log position 1.}

\end{enumerate}




\newpage
\section{Revisiting the FaB View-Change} \label{chap:fab}
\subsection{Introduction}

The Zyzzyva protocol borrows from an earlier work called FaB (Fast Byzantine Consensus)~\cite{MA05,MA06}. FaB introduces a family of Asynchronous Byzantine Agreement (ABA) solutions exhibiting reduced latency when the system is behaving synchronously. In particular, it constructs a parameterized variant for 
$n\ge3f+2t+1$, where $t \leq f$, that has optimal synchronous latency when no more than $t$ non-leader members fail. Putting $t=0$, we obtain a similar setting to Zyzzyva, and the guarantee of a fast execution in fail-free runs. In this paper for simplicity we focus on the case where $n$ is minimal for a given $f$ and $t$ (so $n=3f+2t+1$).

Briefly, the core mechanism for transferring safe values across views revolves around a
``progress certificate''. 
The certificate consists of signed new-view messages from a quorum of $n-f$ replicas
to the leader of a new view. A new-view message from a replica contains the last pre-proposed message accepted by this replica, and the last commit-certificate it received.
A progress certificate is said to ``vouch for'' a value $v$ if it is safe for the leader of the new view to pre-propose $v$.

As we describe below, there is a bug in Parameterized FaB such that the progress certificate may vouch for no value at all, resulting in the protocol getting stuck.


Zyzzyva borrows from FaB the idea of an optimistic fast track, and enhances the approach in a number of dimensions. Zyzzyva provides a state replication protocol, whereas FaB is a single shot consensus solution. Zyzzyva employs speculation in the execution of state updates, allowing a high throughput pipeline of state-machine replication, which is out of the FaB scope. Finally, Zyzzyva includes view-numbers in the view-change protocol, which prevent the ``stuck'' situation in FaB that we expose here.

\subsection{A Skeletal Overview of the FaB Protocol Family}

Martin and Alvisi introduce Fast Byzantine Consensus (FaB) 
in~\cite{MA05, MA06}, a family of protocols parameterized by various resilience assumptions.
The papers use the Paxos terminology to model roles: \textit{proposers}, \textit{acceptors}, and
\textit{learners}. And it employs \textit{proposal numbers} to enumerate proposals. We will adhere to the Zyzzyva (and PBFT) terminology, and translate those to leaders, replicas, and view-numbers. 

FaB has two variants.  The first FaB variant works with $n=5f+1$ replicas, among which leaders are chosen to drive agreement in views. 
We will refer to this variant as FaB5. The second one is parameterized with $n=3f+2t+1$, and we refer to it as PFaB.

\paragraph{$\mathbf{5f+1}$ FaB.}

The basic FaB5 protocol is an easy two-step protocol.
A leader pre-proposes a value to replicas, who each \textit{accept} one value per view and respond with a \emph{prepare} message. A decision is reached in FaB5 when $4f+1$ replicas send a prepare response for it. 
During periods of synchrony, FaB5 is guaranteed to complete through these two easy steps, despite
up to $f$ arbitrary (Byzantine) non-leader failures.

If progress is stalled, replicas elect a new leader and move to a new view.
The core mechanism in FaB5 for transferring safe values across views is a
\textit{progress certificate}.  
A progress-certificate consists of signed new-view messages (REP) from a quorum of $4f+1$ replicas
to the leader of a new view. A new-view message from a replica contains the value in a prepare message sent by this replica.

A progress-certificate is said to \textit{vouch for} a value $v$ if there does not exist
a set of $2f+1$ new-view messages with some identical accepted value $v'$, where $v' \neq v$.

Intuitively, the reason FaB5 is safe is because if a decision is reached in a
view, then $3f+1$ correct replicas prepared it. If the next view is activated, then in every progress-certificate quorum, $2f+1$
of the quorum will prevent vouching for any conflicting proposal. Hence, no correct replica will ever
override an accepted value.

The reason FaB5 is live is because there cannot be two sets of $2f+1$ vouching
against each other's value. Hence, there always exists a safe value to propose.

\paragraph{Parameterized FaB.}

The second FaB variant is called Parameterized FaB (PFaB for short). 
PFaB borrow the idea of an optimistic fast execution track from a long line of works on early-stopping consensus, and in particular, from the optimistic asynchronous Byzantine agreement protocol of Kursawe in~\cite{K02}.

PFaB is parameterized with $n=3f+2t+1$, where $t \le f$. 
It works in two tracks, a
fast track and a recovery track.  The fast track is the same as FaB5, allowing a decision in two
steps if $n-t$ replicas accept a leader proposal. The fast track is guaranteed to
complete in periods of synchrony with a correct leader and up to $t$ Byzantine
replicas. 

Different from FaB5, Parameterized FaB does not necessarily guarantee fast progress
even in periods of synchrony, if the parameter $t$ threshold of failures is exceeded. 
That is, although PFaB is always safe despite up to $f$ Byzantine failures,
it is not always fast.  The fast track is guaranteed to complete during periods of synchrony
in two steps only if the number of actual Byzantine failures does not exceed $t$. 

If progress is stalled, PFaB allows progress via a recovery protocol, which is
essentially PBFT (adapted to $n=3f+2t+1$). 

More precisely, in PFaB, the recovery track revolves around forming a commit-certificate called a \textit{commit-proof}.  When replicas accept a leader proposal, in addition to
sending \textit{prepare} messages ({\tt ACCEPTED}) to the leader, replicas also
send signed prepare messages to each other.  When a replica receives in a view
$(n-f-t)$ prepare messages for the same value, it forms a commit-certificate, and 
sends it in a \textit{commit} message ({\tt COMMITPROOF}) to other replicas.  



A decision is reached if either $n-t$ prepare messages are sent (for the same
value), or $(n-f-t)$ commit messages are sent (for the same value).

As in Fab5, the core mechanism in PFaB for transferring safe values across views is a progress certificate containing new-view messages ({\tt REP}) from a quorum of $n-f$ replicas. Differently, in PFaB, a new-view message from a replica contains both the last value it sent in a prepare message, and the last commit-certificate it sent in a commit message. 

In PFaB, a progress-certificate is said to \textit{vouch for} a value $v$ if there does not exist
a set of $f+t+1$ new-view messages with an identical prepare value $v'$ such that $v' \neq v$; and
there does not exist any commit-certificate with value $v'$ such that $v' \neq v$.

\subsection{Getting Stuck}

In this section, we demonstrate that a progress-certificate may contain
$f+t+1$ new-view messages with some prepare value, and a commit-certificate with a different value. This causes PFaB to get stuck because there is no value vouched-for by the certificate, hence new leaders cannot make any valid proposal. 

For a scenario, we set $f=1$, $t=0$, $n=3f+2t+1=4$.
Denote the replicas by $i_1$, $i_2$, $i_3$, $i_4$, one of whom, say $i_1$, is Byzantine.
The scenario goes through one view change.

\paragraph{View 1:}

  \begin{enumerate}
  \item Leader $i_1$ (Byzantine) pre-proposes value $A$ to $i_2$, $i_3$.

  \item $i_1$, $i_2$, and $i_3$ accept the proposal and send prepare ({\tt ACCEPTED}) messages. Their prepare
messages reach only $i_2$, and $i_2$ forms a commit-certificate ({\tt COMMITPROOF}) for the value $A$.

  \item Meanwhile, the leader $i_1$ equivocates and pre-proposes $B$ to $i_4$. 
  
  \item All further prepare messages other than those sent to $i_1$ are delayed. The delay triggers a view change.
  
  \end{enumerate}

\paragraph{View 2:}

  \begin{enumerate}

  \item The new leader $i_2$ collects a progress certificate consisting of new-view messages ({\tt REP}) from a quorum of $3$ replicas (including itself):

	\begin{itemize}
                \item from $i_1$, the new-view message contains the value $B$, and no commit-certificate.
                \item from $i_2$, the new-view message contains the value $A$, and a commit-certificate for it.
                \item from $i_4$, the new-view message contains the value $B$, and no commit-certificate.
     \end{itemize}
	
  \end{enumerate}

Now we are stuck. This progress certificate contains $2$ messages (from
$i_1$,$i_4$) with prepare value $B$. Hence, the certificate does not vouch for $A$.
At the same time, it contains a commit-certificate (from $i_2$) with value $A$. 
Hence, it does not vouch for $B$ either.

The PFaB paper includes an argument that all process certificates vouch for at least one value (Lemma 7), but unfortunately it has a mistake.






\newpage


\bibliographystyle{plain}
\bibliography{bft}

\end{document}